# Geophysical Classification of Planets, Dwarf Planets, and Moons:

# A Mass Scale and Composition Codes

David G. Russell
Owego Free Academy, Owego, NY USA
russelld@oacsd.org
dgruss23@yahoo.com

## Abstract

A planetary mass scale and system of composition codes are presented for describing the geophysical characteristics of exoplanets and Solar System planets, dwarf planets, and spherical moons. The composition classes characterize the rock, ice, and gas properties of planetary bodies. The planetary mass scale includes five mass classes with upper and lower mass limits derived from recent studies of the exoplanet mass-radius and mass-density relationships, and the physical characteristics of planets, dwarf planets, and spherical moons in the Solar System. The combined mass and composition codes provide a geophysical classification that allows for comparison of the global mass and composition characteristics of exoplanets with the Solar System's planets, dwarf planets and spherical moons. The system is flexible and can be combined with additional codes characterizing other physical, dynamical, or biological characteristics of planets.

## 1. Introduction

To date nearly 3000 exoplanets have been confirmed (Han et al. 2014). It is possible to derive reasonable models for the interior structure of most Solar System planets and many exoplanets (Baraffe et al. 2008, 2010, 2014; Fortney et al. 2006, Sotin et al. 2010). Numerous names are used to characterize exoplanets including: Earth, super-Earth, mini-Neptune, Neptune, sub-Neptune, hot Neptune, Saturn, Jupiter, hot Jupiter, Jovian, gas giant, ice giant, rocky, terrestrial, water, and ocean. There are no clear guidelines for when to apply one of these terms to a given exoplanet. What is the acceptable range in mass, composition, temperature, orbital elements, or other traits that the exoplanet must have in order to be considered an Earth, Neptune, or Jupiter? How much greater or lesser a mass can an exoplanet have before it should be called a 'super-' or 'mini-' version of its Solar System analog? What descriptors should be used for planets with a combination of mass and composition characteristics not found in the Solar system?

As the catalog of confirmed exoplanets continues to grow there is an increasing need for a system of codes to characterize fundamental geophysical and dynamical characteristics of these planets. Plávalová (2012) proposed a classification system that includes codes for exoplanetary mass, semi-major axis, eccentricity, primary composition, and mean Dyson temperature. Other taxonomic systems have been proposed that characterize mass, spectral and atmospheric types, plausibility of life, and stellar type (Irwin & Schulze-Makuch 2001; Sudarsky, Burrows, & Hubeny 2003; Kaelin 2006; Lundock et al. 2009; Encrenaz 2010; Mendez 2011; Fletcher et al. 2014).

In this paper codes are described for classifying the mass and composition characteristics of exoplanets and the planets, dwarf planets, and spherical moons of the Solar System – collectively referred to as *"planetary bodies"*. The codes provide a geophysical taxonomy but can serve as a foundation for additional codes describing other physical, dynamical, and biological characteristics. This paper is organized as follows: In §2 a system of composition codes is described to characterize the rock, ice, and gas properties of planetary bodies. In §3 a planetary mass scale with physically motivated mass boundaries is presented. In §4 the mass scale and composition codes are combined into a geophysical taxonomy useful for characterizing planetary bodies within the Solar System and exoplanetary systems. Concluding remarks are provided in §5.

## 2. Composition codes for characterizing planets, dwarf planets, and moons

### *2.1 General planetary composition classes*

Each planetary body is a unique product resulting from (1) the composition and structure of the gas cloud and the proto-planetary disk from which the planetary system formed, (2) the location, composition, temperature, timescale, and migration history of the body within the proto-planetary disk during formation, and (3) the proto-planetary and planetary interaction and collision history. Planets, dwarf planets, and spherical moons fall into three general composition classes:

(1) ***Rock***: Rock planets are primarily composed of the elements Mg, Si, O, and Fe (Baraffe et al. 2014) in the form of silicate and Fe-rich silicate rocks in a crust and mantle with an iron-rich core.
(2) ***Ice***: The elements carbon, nitrogen, hydrogen, and oxygen form molecules such as $H_2O$, $CH_4$, $NH_3$, CO, $N_2$ and $CO_2$ that are collectively referred to as 'astrophysical' or 'planetary' ices and comprise most of the mass of ice planets (Stern & Levison 2002; Baraffe et al. 2014; de Pater & Lissauer 2015).
(3) ***Gas***: The elements H and He are the primary mass component of gas planets (Baraffe et al. 2010).

For detailed discussions of planetary internal structure see reviews by Baraffe et al. (2010, 2014), Schubert et al. (2010), Sotin et al. (2010), and de Pater & Lissauer (2015). As a population, the planets, dwarf planets, and spherical moons of the Solar System have diverse structure and composition. Models indicate the composition of each planetary body in the Solar System is at least 50% by mass of one of the three composition types (rock, ice, gas), but many of these bodies have a significant mass contribution from a second composition component (Baraffe et al. 2010, 2014; Sotin et al. 2010; de Pater & Lissauer 2015) . For example, Uranus and Neptune are >50% ice by mass but have significant rock and gas fractions. The gas giants Jupiter and Saturn are both > 50% H/He gases but have different Z>2 element mass fractions (Baraffe et al. 2014). The spherical moons of the outer planets have varying fractions of rock and ice. Many sub-Neptune mass exoplanets have no Solar System analog and are >90% rock by mass but with an envelope of H/He gases that significantly inflates the planetary radius (e.g. Lissauer et al. 2013). Therefore, a simple three composition class system (rock, ice, gas) fails to sufficiently characterize the observed variation in the composition and structure of planetary bodies.

**Table 1**. Planet Composition Classes

| Composition codes | Composition name | Composition description |
|---|---|---|
| $R_M$ | Metallic Rock | Silicates & iron with >~50% by mass iron |
| $R_S$ | Silicate Rock | >50% silicate minerals with negligible ices and <50% iron |
| $R_I$ | Icy Rock | >50% by mass by mass silicates/iron with ices <50% & no gas |
| $R_G$ | Rock gas envelope | >50% by mass silicates/iron with an H/He envelope |
| $R_{IG}$ | Rock ice + gas envelope | >50% by mass silicates/iron with ices and an H/He envelope |
| $R_?$ | Rock (?) | >50% by mass silicates/iron with uncertain fractions of ice & gas |
| I | Ice | > 90% ices by mass, < 10% silicates & iron, no H/He envelope |
| $I_S$ | Rock Ice | >50% ices by mass with >10% silicate & iron, no H/He envelope |
| $I_{SG}$ | Ice giant | >50% ices by mass w/rock core & H/He envelope |
| $I_G$ | Ice Gas | >50% ices by mass w/ice core & H/He envelope |
| $I_?$ | Ice (?) | >50% ices by mass with highly uncertain fractions of rock & gas |
| $G_{01}$ | Gas | >50% H/He gas with 0-10% Z>2 elements |
| $G_{12}$ | Gas | >50% H/He gas with 10-20% Z>2 elements |
| $G_{23}$ | Gas | >50% H/He gas with 20-30% Z>2 elements |
| $G_{02}$ | Gas | >50% H/He gas with 0-20% Z>2 elements |
| $G_{13}$ | Gas | >50% H/He gas with 10-30% Z>2 elements |
| $G_{24}$ | Gas | >50% H/He gas with 20-40% Z>2 elements |
| $G_D$ | Gas Deuterium Burner | >50% H/He gas with deuterium burning |
| $G_{05}$ | Gas (?) | >50% H/He gas with 0-49% Z>2 elements |

The composition codes are summarized in Table 1 and described in §2.2 - 2.5. The composition codes for planets, dwarf planets and spherical moons begin with "R", "I", and "G" for planetary bodies that are >50% by mass rock, ice, and gas respectively. Further resolution of composition is characterized with subscripted letters or numbers that follow R, I, and G. The composition codes are applied to the planets, dwarf planets and spherical moons of the Solar System in Table 2.

## *2.2 Rocky planetary bodies: $R_S$, $R_M$, $R_I$, $R_G$, $R_{IG}$*

The Solar System's rocky planetary bodies include (1) terrestrial planets and several spherical moons with pure rock composition and (2) dwarf planets and icy moons with >50% rock and a significant fraction of astrophysical ices. The known exoplanet population also includes rocky planets with radii inflated by an H/He or H/He/$H_2O$ envelope that accounts for <10% of the planetary mass.

### *2.2.1 $R_S$ (silicate) planetary bodies*

Planetary bodies with $R_S$ class have pure rock composition with >50% by mass silicate rocks and <50% Fe. Solar System $R_S$ bodies include Venus, Earth, Mars, the Moon, Io, and Vesta. The composition of Venus, Earth, and Mars is ~65-75% silicate rock and ~25-35% Fe (Morgan & Anders 1980; Ronco et al. 2015). The structure of $R_S$ planets include a solid and/or liquid core composed of Fe-Ni metal, Fe, or FeS, a silicate mantle, and an outer silicate crust (Sotin et al. 2010; Baraffe et al. 2014; de Pater & Lissauer 2015). Based upon the mass-radius relationship, examples of rocky exoplanets with $R_S$ classification include Kepler 10b (Dumusque et al. 2014), Kepler 36b (Carter et al. 2012), Kepler 78b (Grunblatt et al. 2015), and Kepler 93b (Ballard et al. 2014; Dressing et al. 2015).

### *2.2.2 $R_M$ (metallic) planetary bodies*

Planetary bodies with $R_M$ class are "metallic" planets having pure rock composition but with >50% by mass Fe. Mercury is the only $R_M$ planet in the Solar System and has an iron core that accounts for ~64% of the planetary mass and ~75% of the planetary radius surrounded by a silicate mantle and crust (de Pater & Lissauer 2015). The exoplanet CoRoT 7-b (5.74 $M_\oplus$; $\rho$ = 7.5 g $cm^{-3}$) is a candidate $R_M$ composition class planet (Wagner et al. 2012; Barros et al. 2014).

### *2.2.3 $R_I$ (icy rock) planetary bodies*

"$R_I$" planetary bodies have >50% rock by mass but contain a significant fraction of astrophysical ices. Several examples of $R_I$ planetary bodies in the Solar System include the moons Ganymede, Titan, Triton, Europa and dwarf planets such as Ceres, Pluto, and Eris. The icy rock planetary bodies found in the Solar System have an interior structure with a rock or iron core overlain by a silicate mantle, most likely at some depth an interior liquid water layer, and an icy crust (McKinnon et al. 2008; de Pater & Lissauer 2015). In some instances the mantle may include an icy mantle layer overlaying an interior rock mantle (Ganymede) or an icy mantle directly overlaying the core (Triton).

Exoplanet candidates for $R_I$ composition class are super-Earth to Neptune mass planets including Kepler 10c (Dumusque et al. 2014; Weiss et al. 2016), Kepler 68b (Gilliland et al. 2013), HD97658b (Van Grootel et al. 2014), and Kepler 18b (Cochran et al. 2011). Some $R_I$ class super-Earth planets may represent very massive versions of Ganymede (Sotin et al. 2010). The structure of massive $R_I$ composition class exoplanets will include a massive rock core surrounded by an $H_2O$-rich envelope or under the right circumstances the body may be an ocean planet with a 50 km to 475 km ocean overlaying high pressure ice and a rock core (Sotin et al. 2010).

### *2.2.4 $R_G$ and $R_{IG}$ planetary bodies*

Exoplanet surveys have identified a class of super-Earth and sub-Neptune planets that are most likely rock planets with a radius inflated from a <10% by mass H/He envelope rather than astrophysical ices. These planets will have a rock core containing over 90% of the planetary mass surrounded by an H/He envelope. Planets of this composition type have the code "$R_G$". Examples of candidate exoplanets for the $R_G$ class include Kepler 11c-f (Lissauer et al. 2013) and Kepler 20c (Gautier III et al. 2012).

GJ 436b (Gillon et al. 2007; Nettelmann et al. 2010) and HD97658b (Van Grootel et al. 2014) are exoplanets for which available data makes it hard to rule out an $R_{IG}$ composition that is >50% rock but with an envelope that is composed of both astrophysical ices and H/He gas.

### *2.3 Icy planetary bodies: I, $I_S$, $I_{SG}$, $I_G$*

Planetary bodies in the Solar System with >50% of the mass comprised of astrophysical ices include the ice giants (Uranus & Neptune) and numerous spherical moons of the outer planets (Table 2).

#### *2.3.1 I (Ice) planetary bodies*

Planetary bodies of the "I" composition class have a composition that is >90% astrophysical ices by mass and are therefore nearly "pure" ice. Saturn's moon Tethys ($\rho$ = 0.97 g $cm^{-3}$) is the only known planetary body in the Solar System with an almost pure ice composition (de Pater & Lissauer 2015).

#### *2.3.2 $I_S$ (rocky ice) planetary bodies*

The Solar System's gas and ice giant planets have numerous spherical moons that are icy bodies with an increasing fraction of rock toward the center overlain by an icy mantle, a possible interior liquid $H_2O$ layer, and an ice crust. Bodies with this composition are given the code "$I_S$" and include the moons Callisto, Iapetus, Mimas, and others listed in Table 2. Candidate $I_S$ exoplanets include Kepler llb (Lissauer et al. 2013), Kepler 68b (Gilliland et al. 2013), Kepler 18b (Cochran et al. 2011), and Kepler 20b and 20c (Gautier III et al. 2012).

#### *2.3.3 $I_{SG}$ (ice giant planets)*

Uranus and Neptune represent the class of planets identified as "ice giants" and have a composition that is ~60-65% ices, ~25% rock and ~10-15% H and He (Baraffe et al. 2010). Models of the interior structure of the ice giants include a rocky core, a liquid ionic icy mantle, and a molecular $H_2$, He, and $CH_4$ envelope (Baraffe et al. 2010, 2014; de Pater & Lissauer 2015). The composition code for ice giants is "$I_{SG}$" indicating an ice planet with significant mass contributions from both silicates and He/He gases. Note that ice giants may be similar to $R_{IG}$ planets that also have significant mass contributions from rock, ice, and gas but ice giants are > 50% ice by mass whereas $R_{IG}$ planets are >50% rock by mass.

Exoplanets with possible ice giant composition include Kepler 18c and 18d (Chochran et al. 2011), GJ 3470b (Demory et al. 2013), and Kepler 101b (Bonomo et al. 2014).

### *2.3.4 $I_G$ planets*

Planets with an $I_G$ composition class will have >50% by mass planetary ices with most of the remaining mass in the form of an H/He envelope. An interesting exoplanet candidate for this planet class is the extremely low density planet Kepler 87c which could also have an ice giant composition ($I_{SG}$), or an $R_G$ composition (Ofir et al. 2014).

## ***2.4 Gas (H/He) planetary bodies: $G_Z$, $G_D$***

### *2.4.1 $G_Z$ (gas giant planets)*

According to the core accretion model gas giant planets form by initial accretion of an approximately 10 $M_\oplus$ rocky core upon which a massive H/He envelope is collected from the proto-planetary disk (Baraffe et al. 2010; Helled et al. 2014). The internal structure of a gas giant will include a rock and ice core overlain by a convective liquid metallic hydrogen and helium envelope and an outer molecular hydrogen and helium envelope (de Pater & Lissauer 2015). The mass fraction of Z>2 elements in gas giants can show considerable variation. Structural models for Jupiter and Saturn indicate Z>2 element mass fractions of ~0.08 and ~0.22 respectively (Baraffe et al. 2010). Kepler 30d is a gas planet with a Z>2 mass fraction possibly as large as 0.40 (Spiegel, et al. 2014). The composition code for gas giant planets therefore includes subscripted numbers indicating a range for the mass fraction of Z>2 elements. Jupiter is a $G_{01}$ planet where the subscript "01" indicates that the mass fraction of Z>2 elements is between 0.00 and 0.10. Saturn is a $G_{23}$ planet indicating a Z>2 element mass fraction between 0.20 and 0.30.

Examples of exoplanets with a $G_{01}$ composition class include KOI 680b (Almenara et al. 2015) and Kepler 423b (Gandolfi et al. 2015). The exoplanets KOI 614b (Almenara et al. 2015) and Kepler 77b (Gandolfi et al. 2013) have a Z>2 element fraction between 0.0 and 0.20 and therefore have a $G_{02}$ composition code. When a gas giant exoplanet has a highly uncertain fraction of Z>2 elements it is given the composition code $G_{05}$.

### *2.4.2 $G_D$ (deuterium burning gas giants)*

It has been suggested that giant planets formed by core-accretion in a proto-planetary disk can exceed the deuterium burning (DB) limit of 13 Jupiter masses (Mulders et al. 2013; Baraffe et al. 2014; Ma & Ge 2014) and therefore can be classified as 'deuterium burning planets' (Baraffe et al. 2008; Molliere & Mordasani 2012). These giant planets are given the composition code "$G_D$". KOI-423b (18 $M_J$) is candidate $G_D$ exoplanet (Bouchy et al. 2011).

**Table 2**. Planets, dwarf planets and spherical moons of the Solar System

| Planetary Body | Mass (kg)* | Radius (km)* | Density g cm$^{-3}$* | Composition class |
|---|---|---|---|---|
| Jupiter | 1.90 x $10^{27}$ | 69911 | 1.326 | $G_{01}$ c,d |
| Saturn | 5.68 x $10^{26}$ | 58232 | 0.687 | $G_{23}$ c,d |
| Neptune | 1.02 x $10^{26}$ | 24622 | 1.638 | $I_{SG}$ c,d |
| Uranus | 8.68 x $10^{25}$ | 25631 | 1.270 | $I_{SG}$ c,d |
| Earth | 5.97 x $10^{24}$ | 6371 | 5.513 | $R_S$ c,d |
| Venus | 4.87 x $10^{24}$ | 6052 | 5.243 | $R_S$ c,d |
| Mars | 6.40 x $10^{23}$ | 3390 | 3.934 | $R_S$ c,d |
| Mercury | 3.30 x $10^{23}$ | 2440 | 5.427 | $R_M$ c,d |
| Ganymede | 1.48 x $10^{23}$ | 2631 | 1.94 | $R_I$ c |
| Titan | 1.30 x $10^{23}$ | 2575 | 1.88 | $R_I$ c,e |
| Callisto | 1.08 x $10^{23}$ | 2410 | 1.83 | $I_S$ c |
| Io | 8.93 x $10^{22}$ | 1822 | 3.53 | $R_S$ c |
| Moon | 7.35 x $10^{22}$ | 1738 | 3.34 | $R_S$ c |
| Europa | 4.80 x $10^{22}$ | 1561 | 3.01 | $R_I$ c |
| Triton | 2.10 x $10^{22}$ | 1353 | 2.06 | $R_I$ c,e |
| Eris | 1.67 x $10^{22}$ | 1163 | 2.50 | $R_I$ |
| Pluto | 1.31 x $10^{22}$ | 1151 | 2.05 | $R_I$ e |
| Titania | 3.4 x $10^{21}$ | 789 | 1.66 | $I_S$ |
| Oberon | 2.9 x $10^{21}$ | 761 | 1.56 | $I_S$ |
| Rhea | 2.3 x $10^{21}$ | 764 | 1.23 | $I_S$ e |
| Iapetus | 1.8 x $10^{21}$ | 736 | 1.08 | $I_S$ |
| Charon | 1.55 x $10^{21}$ | 604 | 1.68 | $R_I$ e |
| Ariel | 1.3 x $10^{21}$ | 599 | 1.59 | $I_S$ |
| Umbriel | 1.2 x $10^{21}$ | 585 | 1.46 | $I_S$ |
| Dione | 1.1 x $10^{21}$ | 562 | 1.48 | $I_S$ |
| Ceres | 9.35 x $10^{20}$ a | 476 | 2.14 b | $R_I$ |
| Tethys | 6.18 x $10^{20}$ | 533 | 0.97 | I e |
| Vesta | 2.71 x $10^{20}$ a | 265 | 3.7 b | $R_S$ f |
| Pallas | 2.41 x $10^{20}$ a | 273 | 3.2 b | $R_I$ |
| Enceladus | 1.08 x $10^{20}$ | 252 | 1.61 | $R_I$ e |
| Miranda | 6.6 x $10^{19}$ | 236 | 1.21 | $I_S$ |
| Proteus | 5.0 x $10^{19}$ | 210 | 1.3 | $I_S$ |
| Mimas | 3.75 x $10^{19}$ | 198 | 1.15 | $I_S$ e |

Notes:
*Planetary data is from the NASA Solar System Exploration webpage available at http://solarsystem.nasa.gov/planets/index.cfm

a – Michalak (2001)
b – Michalak (2000)
c - de Pater & Lissauer (2015)
d – Baraffe et al. (2010, 2014)
e – Schubert et al. (2010)
f – Russell et al. (2012)

### *2.5 Advantages of the composition codes*

There are several advantages to the system of composition codes described above. The system is relatively simple as R, I, and G indicate planets with > 50% by mass rock, ice, or gas respectively. Subscript modifiers then allow higher resolution description of planetary composition within the uncertainty of composition and interior structure models. The system allows exoplanets similar to Solar System planets to be identified but also includes codes for exoplanets that have no Solar System analogs. For example, the Kepler 11 planetary system may contain four planets with $R_G$ composition (Lissauer et al. 2013). The rock mass fraction in $R_G$ exoplanets is >0.9 but these planets are not typical of terrestrial planets as the radius is inflated by an H/He envelope with a mass fraction that is <0.1 of the planet's mass. In the absence of a set of composition codes, characterization of $R_G$ planets is difficult. Kepler 10c is a 17 +/-2 $M_\oplus$ planet with an H/He envelope fraction of only 0.5% or a more massive water envelope (Lopez & Fortney 2014). Wolfgang & Lopez (2015) prefer to characterize Kepler 10c as a sub-Neptune rather than a super-Earth. However, $R_G$ planets have a dominant rock composition whereas Neptune has a dominant ice composition. Earth-like terrestrial planets lack the H/He envelope found in $R_G$ planets. Therefore, the names "super-Earth" and "sub-Neptune" both fail to accurately describe the composition of Kepler 10c whereas the composition code $R_G$ characterizes the planet's composition while also identifying the composition differences from the Earth and Neptune.

The composition codes can be related to other composition descriptions. For example, Grasset et al. (2009) identified four composition classes of super-Earth planets: Mercury-like (iron-rich), Earth-like (silicate-rich), water-rich, and Neptune-like that can be distinguished with mass-radius relationships. The composition codes presented in this section clearly classify each of these super-Earth types: $R_M$ (Mercury-like), $R_S$ (Earth-like), $R_I$ (water-rich), and $I_{SG}$ (Neptune-like).

## 3. A Planetary Mass Scale

The Solar System's planets, dwarf planets, and spherical moons range in mass from icy moons such as Mimas, Proteus, and Miranda with masses on the order of $10^{19}$ kg to the gas giant Jupiter with a mass of $10^{27}$ kg. Most exoplanets and brown dwarfs discovered to date include bodies ranging from ~1 Earth mass ($M_\oplus$) to ~60 Jupiter mass ($M_J$). Mass-density and mass-radius relationships indicate several mass ranges that broadly differentiate planetary composition types. Gas giant planets generally have masses of ~0.3 to ~60 $M_J$ (Hatzes & Rauer 2015). Most terrestrial rock planets have radii <1.6 $R_\oplus$ and masses <6 $M_\oplus$ (Rogers 2015; Weiss & Marcy 2014; Marcy et al. 2014). The population of planets between 6 and 60 $M_\oplus$ has varying fractions of rock, ice, and gas including ice giants ($I_{SG}$), $R_G$, $R_I$, and $I_S$ planets (Grasset et al. 2009; Marcy et al. 2014).

At the high mass end of sub-stellar bodies (>1 $M_J$) the distinction between giant gas planets and brown dwarfs is complicated by the incompatibility between deuterium burning and formation based definitions, observations of the mass distribution of planets and brown dwarfs, and the observed mass-density relationship. The IAU has defined brown dwarfs as objects that exceed the deuterium burning (DB) limit (~13 $M_J$) irrespective of formation mechanism. As defined by formation theories giant planets (GP) are objects formed in a proto-planetary disk around a star by core accretion or disk instability mechanisms whereas brown dwarfs (BD) are objects formed in a star-like manner from gas fragmentation and collapse in molecular clouds (see reviews by Kumar 2003; Chabrier et al. 2007, 2014; Whitworth et al. 2007; Luhman 2008; Baraffe et al. 2010; D'Angelo et al. 2010). However the mass functions for objects that form in a protoplanetary disk and those that form by gas collapse overlap with gas collapse mechanisms forming bodies as small as 5-6 $M_J$ (Zapatero Osorio et al. 2000, 2002; Lucas et al. 2006; Caballero et al. 2007; Bihain et al. 2009; Delorme et al. 2012; Peña Ramírez et al. 2012; Leconte et al. 2009; Scholz et al. 2012; Luhman 2014; Beichman et al. 2014; Brandt et al. 2014) while planets formed in a proto-planetary disk can exceed the DB limit (Mulders et al. 2013; Baraffe et al. 2014; Ma & Ge 2014).

The BD-GP overlapping mass regime makes it problematic to simultaneously apply the DB limit and the formation mechanism as criteria for distinguishing GP and BD (Chabrier et al. 2014). The mass-density and mass-radius relationships from 0.3 to 60 $M_J$ lack any distinguishing features that identify a mass boundary separating the GP and BD populations (Hatzes & Rauer 2015; Chen & Kipping 2016). Instead, the mass-density relationship indicates that all objects in the mass range 0.3 to 60 MJ have the same physics controlling the structure of this population of sub-stellar bodies (Hatzes & Rauer 2015).

Chabrier et al. (2014) has suggested that formation mechanism should be used to define the difference between GP and BD noting that objects that form in a proto-planetary disk will have Z>2 element enrichment relative to the parent star and therefore are distinct objects from those formed by gas collapse mechanisms (Whitworth et al. 2007; Baraffe et al. 2010; Chabrier et al. 2014). Planets formed in a proto-planetary disk around a star or BD that exceed the DB limit are then identified as 'deuterium burning planets' (Baraffe et al. 2008; Molliere & Mordasani 2012).

The planetary mass scale described in the following sections takes into account the observations described above. The mass codes P1, P2, P3, P4, and P5 have physically motivated mass boundaries derived from mass-radius and mass-density relationships along with the characteristics of Solar System planets and exoplanets. Within each mass class the planet population will have a much narrower range of composition and structure relative to the full range of composition types among all known planets. With the exception of the P1 mass class, each class in Table 3 is named after the most massive planetary body found in the Solar System within the mass class. The names do not require a composition similar to the planet after which the mass class is named.

**Table 3**. Planetary Mass Scale

| Class | Class Name | Mass (Kg) | Mass (J) | Mass (⊕) |
|---|---|---|---|---|
| P1 | Brown Dwarf Mass | $9.5 \times 10^{27} – 1.2 \times 10^{29}$ | 5 - 60 | |
| P2 | Jupiter Mass | $3.6 \times 10^{26} - 9.5 \times 10^{27}$ | 0.2 – 5 | 60 – 1600 |
| P3 | Neptune Mass | $3.6 \times 10^{25} – 3.6 \times 10^{26}$ | | 6 - 60 |
| P4 | Earth Mass | $3.0 \times 10^{23} - 3.6 \times 10^{25}$ | | .05 - 6 |
| P5 | Ganymede Mass | $3.7 \times 10^{19} - 3.0 \times 10^{23}$ | | |

### *3.1 P1 Class (Brown dwarf mass)*

The mass range for the P1 mass class is 5-60 $M_J$ and includes the most massive gas giant planets formed in a proto-planetary disc and overlaps the mass range of brown dwarfs formed by gas collapse. The upper mass limit for the P1 mass class is 60 $M_J$ and is the upper limit for sub-stellar objects as indicated by the mass-density relationship (Hatzes & Rauer 2015). The lower mass limit for the P1 mass class is 5 $M_J$ – the lower mass limit for bodies formed by gas collapse mechanisms (Lucas et al. 2006; Caballero et al. 2007; Delorme et al. 2012; Peña Ramírez et al. 2012; Scholz et al. 2012; Brandt et al. 2014, Luhman 2014; Beichman et al. 2014).

While the lower mass limit for the P1 class is approximately the smallest mass for objects that can form by gas collapse, planets with masses larger than 5 $M_J$ may form by core accretion or disk instability in a proto-planetary disk (LeConte et al. 2009; Mulders et al. 2013; Chabrier et al. 2014, Ma & Ge 2014) and therefore the population of bodies with masses >5 $M_J$ includes both GP and BD as defined by formation mechanisms. Most sub-stellar bodies with masses <5 $M_J$ will have formed in a proto-planetary disk and have Z>2 element enrichment relative to the parent star or BD (see reviews by Whitworth et al. 2007; Chabrier et al. 2014).

### *3.2 P2 Class (Jupiter mass)*

Planets in the P2 class have masses ranging from 0.02 $M_J$ to 5 $M_J$ or 60 $M_\oplus$ to 1600 $M_\oplus$. The upper mass limit for the P2 class is the lower mass limit for objects that reside in the BD-GP overlapping mass regime (§ 3.1). The lower mass limit for the P2 class is approximately the mass limit where the population of planets transitions from bodies having a >50% by mass H/He envelope to less massive planets with a higher percentage of planetary ices and rock (Lopez & Fortney 2014). The mass-density relationship demonstrates that the transition between gas giants and planets with a dominant ice/rock composition occurs somewhere between 0.1 $M_J$ and 0.3 $M_J$ (Hatzes & Rauer 2015; Laughlin & Lissauer 2015). Therefore 0.2 $M_J$ – or 60 $M_\oplus$ - is adopted as the lower mass limit for the P2 mass class. Most P2 planets should have a gas giant composition similar to Jupiter and Saturn.

### *3.3 P3 Class (Neptune Mass)*

Planets in the P3 class have masses ranging from 6 $M_\oplus$ to 60 $M_\oplus$ with the justification for the upper mass boundary described in § 3.2. The lower mass limit for the P3 mass class is 6 $M_\oplus$. Mass-radius relationships indicate that exoplanets with radii < ~1.6 $R_\oplus$ or a mass <~ 6 $M_\oplus$ are primarily of terrestrial composition (Marcy et al. 2014; Weiss & Marcy 2014; Buchhave et al. 2014; Howe, Burrows & Verne 2014; Lopez & Fortney 2014; Rogers 2015; Dressing et al. 2015). Therefore the lower mass limit for the P3 class separates the population of ice giants and rocky planets with H/He or ice envelopes from the population of predominately terrestrial planets. Planets in the P3 class may be true ice giants such as Uranus, Neptune, GJ436b (Gillon et al. 2007), and Kepler 101b (Bonomo et al. 2014). However, planets in the P3 class mass range - often identified as 'Neptunes' - have tremendous variation in composition and in some cases may have a significant fraction of H/He gas (Spiegel, Fortney, & Sotin 2014). For example, Kepler 30d (mass = 23.1 $M_\oplus$, radius = 8.8 $R_\oplus$) is a Neptune mass planet that is only 30 per cent Z>2 elements and 70 per cent H and He gas (Batygin & Stevenson 2013; Spiegel, Fortney, & Sotin 2014).

Exoplanets such as Kepler 30d illustrate that the P3 class name 'Neptune mass' is a different meaning than simply 'Neptune' – which implies a similarity in composition and structure with Neptune. While P3 planets have a Neptune range mass, the P3 name 'Neptune mass' does not require a Neptune composition and structure. Kepler 30d is more accurately characterized as a 'mini-Jupiter' since the mass fraction of H/He is >0.50.

### *3.4 P4 Class (Earth Mass)*

Planets in the P4 mass class have masses ranging from 0.05 $M_\oplus$ to 6 $M_\oplus$. Based upon the mass-radius relationship most planets with radius <~1.6 $R_\oplus$ should be terrestrial planets (Buchhave et al. 2014; Marcy et al. 2014; Weiss & Marcy 2014; Howe, Burrows & Verne 2014; Dressing et al. 2015; Rogers 2015). Rogers (2015) found that most exoplanets with radii >1.6 $R_\oplus$ are too low in density to be pure rock composition and find that > 50% of planets with radius < 1.6 $R_\oplus$ – or ~6 $M_\oplus$ – are "Earth-like".

The lower mass limit for the P4 planet class is 0.05 $M_\oplus$. Among the Solar System's dwarf planets and spherical moons less massive than Mercury only Io, the Moon, and Vesta have a pure rock composition (Table 2). The remaining bodies have $R_I$, $I_S$, or I compositions indicating a significant mass percentage of ices. Ganymede is the largest $R_I$ class planetary body in the Solar System with a mass that is 44.8 % the mass of Mercury. This indicates that within the Solar System the mass range for the population of bodies with a high frequency of pure rock composition only extends to the mass of Mercury (0.055 $M_\oplus$) and therefore 0.05 $M_\oplus$ is adopted as the lower mass limit for the "Earth-like" P4 class planets.

### *3.5 P5 Class (Ganymede mass)*

Planetary bodies in the P5 mass class have masses in the range 3.7 x $10^{19}$ – 3.0 x $10^{23}$ kg. P5 bodies include the Solar System's dwarf planets and spherical moons with most having mixed rock-ice composition classes $R_I$ and $I_S$. The lower mass limit for the P5 mass class is 3.7 x $10^{19}$ kg - the mass of Mimas and approximately the minimum mass required for an icy body formed in a proto-planetary disk to attain a nearly spherical hydrostatic equilibrium shape (Lineweaver & Norman 2010; Tancredi 2010).

**Table 4.** "Super-" and "mini-" Planet classes

| Mass Code | Primary composition (name) | Alternate compositions (names) |
|---|---|---|
| P1 | $G_Z$ (super-Jupiter) | $G_D$ (deuterium burning planet) |
| P2 | $G_Z$ (Jupiter) | $I_{SG}$ (super-Neptune) |
| P3 | $I_{SG}$ (Neptune) | $R_S$, $R_M$ (super-Earth); $G_Z$ (mini-Jupiter) |
| P4 | $R_S$ or $R_M$ (Earth) | $R_I$ or $I_S$(super-Ganymede) |
| P5 | $R_I$ (Ganymede) or $I_S$ | $R_S$ or $R_M$ (mini-Earth) |

### *3.6 'Super-' and 'mini-' planets*

A difficulty with characterizing exoplanets as Earth's, Neptune's, or Jupiter's is that these names indicate not only a general mass range, but are also associated with composition and structural characteristics. However, a 10-20 $M_\oplus$ planet might be predominantly planetary ices by mass and have a Neptune-like ice giant composition and structure or it could be predominantly rock composition. For example Kepler 18c (17.3 $M_\oplus$) and 18d (16.4 $M_\oplus$) appear to have ice giant composition (Cochran et al. 2011) whereas Kepler 10c (13.98 $M_\oplus$ - Weiss et al. 2016) and BD+20594b (16.3 $M_\oplus$ - Espinoza et al. 2016) have smaller radii consistent with >50% rock composition classes $R_S$, $R_G$, and $R_I$. There are also no clear guidelines for how closely the mass of an exoplanet must match its Solar System analog before the planet should be identified as a 'super-' or 'mini-' version of Earth, Neptune, or Jupiter.

The planetary mass scale provides a physically motivated guideline for when to use terms such as 'super-Earth', 'super-Neptune', or 'mini-Jupiter'. The mass limits for the P2, P3, P4, and P5 mass classes indicate mass ranges for planets and planetary bodies that are most likely to be similar to Jupiter, Neptune, Earth, and Ganymede respectively. The prefixes 'super-' and 'mini-' can therefore be applied to planets and planetary bodies with composition that differs from the norm for the mass ranges of the planetary mass codes (Table 4). If a rock planet has a P4 mass class then it is simply an 'Earth' – even if it is 2-6 times the Earth's mass. A rock planet with a P3 mass class – such as CoRoT-7b (Barros et al. 2014) can be identified as a 'super-Earth' since the planetary mass falls above the normal range for the population of planets that are terrestrial rock planets. Likewise a planet such as HD 149026b which has a Jupiter range mass but a Neptune-like composition (Fortney et al. 2006) should be identified as a 'super-Neptune'. Within the Solar System Io could be identified as a "mini-Earth" because it has the general composition and interior structural characteristics of P4 class pure rock planets but a mass in the range for the P5 icy rock planets.

### *3.7 Discussion and comparison with other proposed mass scales*

Mass scales and codes have also been proposed by Kaelin (2006), Mendez (2011), and Plávalová (2012), Chen & Kipping (2017). Kaelin (2006) proposed the K-scale based upon planetary mass-density relations. The most massive planetary bodies in the K-scale are given the class M7 and the least massive are given the class M3. K-scale mass boundaries use a $10^2$ mass difference between the upper limits of each successive mass class and therefore are not physically motivated. For example the upper mass limit for the "Terran" mass class of the K-scale is 1.7 $M_\oplus$ whereas exoplanet studies indicate that the upper mass limit for the terrestrial planet population is ~6 M⊕ (Marcy et al. 2014; Weiss & Marcy 2014; Howe, Burrows & Verne 2014; Dressing et al. 2015; Rogers 2015). The mass scale presented in this paper adopts 6 M⊕ as a physically motivated upper mass limit derived from the mass-radius relationship for the P4 mass class. This mass limit separates planets that are primarily terrestrial from planets in the P3 mass class that have ice giant composition or have radii inflated by a significant mass contribution from ices or H/He gas.

Mendez (2011) has proposed a mass scale that has no codes, but instead uses the names Mercurian, Subterran, Terran, Superterran, Neptunian, and Jovian to identify six mass ranges for planetary bodies. The physical criteria for the six mass ranges are undefined but the boundaries in some cases closely match the mass scale presented in this paper. For example, the upper mass limit of the "Neptunian" mass class is 50 M⊕ - close to the 60 M⊕ upper mass limit for the P3 mass class. However, the upper mass boundary for the "Terran" mass class in the Mendez (2011) proposal is 2 M⊕ which leaves out a significant portion of the terrestrial planet population as determined from the mass-radius relationship (Rogers 2015).

Plávalová (2012) proposed expressing exoplanetary masses in Mercury, Earth, Neptune, and Jupiter mass units. The codes proposed are not tied to any physical criteria and provide a direct mass of the planet in the specified planetary units. While Plávalová has specified exclusive mass ranges for using each Solar System mass unit, the ranges prescribed have no physical basis and may create unnecessary confusion. For example the mass of the ice giant Uranus must be expressed in Earth units while the masses of the ice giant Neptune and the gas giant Saturn are expressed in Neptune mass units. Venus is prescribed to be expressed as 15 Mercury mass units rather than 0.81 Earth mass units. The mass codes summarized in Table 3 have specified kilogram mass ranges but may also be expressed in Earth masses and Jupiter masses when those units are found more useful.

In contrast to the mass scales and codes proposed by Kaelin (2006), Mendez (2011), and Plávalová (2012), the boundaries of the mass scale described in this paper are based upon physical criteria adopted from observed mass-radius, mass-density relationships, and the characteristics of Solar System planets and exoplanets (§3.1-3.5). The resulting mass codes group planets into mass ranges within which the planet population has a narrower range of composition and structural characteristics. For example, most P2 planets will be gas giants, most P3 planets will be ice or rock dominated with an H/He envelope, and most P4 planets will be rock planets. These mass ranges result from physical principles and formation mechanisms. When considering planet masses below ~6 M⊕ it becomes increasingly difficult for a planet to maintain a gas envelope and therefore H/He rich planets should be rare among P4 planets. Likewise, with planetary formation mechanisms such as core accretion it is difficult for planets to accumulate masses in excess of 60 M⊕ without also accreting a massive H/He envelope and therefore P2 rock planets should be extremely rare. However, the mass scale is still compatible with the actual variation of planetary composition that exists within each mass class as the mass code alone does not require a planet have the composition that is most common for the mass range.

Another advantage of the mass scale presented here over the mass scales of Kaelin (2006), Mendez (2011), and Plávalová (2012) is that the use of physical criteria to identify the boundaries of the mass scale also naturally provides a useful guideline for when to use terms such as “mini-Earth”, “super-Earth”, “super-Neptune”, and mini-Jupiter (§3.6).

Based upon the mass-radius relationship Chen & Kipping (2017) have proposed a physically motivated mass scale that includes three planetary mass ranges: Terran worlds (<2M⊕), Neptunian worlds (2M⊕ - 0.41 $M_J$), Jovian worlds (0.41 $M_J$ - ~80 $M_J$). The Jovian worlds of the Chen & Kipping (2017) mass scale approximately covers the full mass range for gas giant planets and incorporates both the P1 and P2 mass classes presented here. While Chen & Kipping (2017) present a simpler system with fewer mass classes the additional mass classes presented in this paper are important. The P1/P2 mass boundary corresponds with the minimum mass for bodies that can form by gas collapse mechanisms and therefore separates lower mass gas giant planets that have formed in a protoplanetary disk (P2 planets) from the higher mass P1 population of sub-stellar bodies that may be gas giant planets or may be brown dwarfs depending upon formation mechanisms and how the difference between GP and BD is defined.

The low mass end of the mass scale proposed by Chen & Kipping (2017) combines all P4 and P5 planets, dwarf planets, and moon into a single “Terran worlds” group. In this paper the P4 and P5 mass classes divide the Solar System’s “Terran worlds” into the pure rock planets with $R_S$ and $R_M$ composition classes (P4) and the dwarf planets and moons that primarily have $R_I$ and $I_S$ composition classes (P5).

## 4. Combined codes for geophysical classification of planetary bodies

The composition and mass codes presented in § 2 and § 3 may be combined to describe the geophysical characteristics of a planet, dwarf planet, or moon. This taxonomy is applied to Solar System planets, dwarf planets, and moons in Table 5 and a sample of exoplanets in Table 6. The first code is the P1 to P5 code from the planetary mass scale followed by the planetary composition code. In this taxonomy the geophysical classification of the Earth is P4$R_S$, of Neptune is P3$I_{SG}$, and of Jupiter is P2$G_{01}$.

The composition and interior structure of exoplanets is often hard to narrow down to a single composition class as multiple combinations of rock, ice, and gas can match the data available for modeling the planetary structure and composition (see Spiegel, Fortney, & Sotin 2014). For example, Kepler 68b has an uncertain composition and models with nearly 50 per cent water or nearly 50 per cent outgassed H/He are equally compatible with the mass, radius, and density for this planet (Gilliland et al. 2013). Therefore many of the exoplanets listed in Table 6 have several possible composition classes listed with the mass code.

Despite the greater overall uncertainty in composition and interior structure for exoplanets, the taxonomy allows for useful comparisons of exoplanets with Solar System planets. There is a significant diversity among the Neptune mass (P3) planets. For example, GJ 436b (Gillon et al. 2007) and Kepler 101b (Bonomo et al. 2014) are ice giants similar to Uranus and Neptune (Gillon et al. 2007). Kepler 18c and 18d have a dominant ice composition but a significant (~20 and 40 per cent by mass respectively) H/He envelope (Cochran et al. 2011) whereas Uranus and Neptune both have ~12% of the planetary mass in an outer H/He envelope (Baraffe et al. 2010). CoRoT-7b is likely a P3 rock planet and is a “super-Earth” or “super-Ganymede”. The population of P3 planets includes $R_G$ planets with no Solar System analog. These planets are >90% by mass rock composition with an H/He envelope accounting for <10% of the planetary mass but a significant fraction of the planetary radius (Lissauer et al. 2013).

Kepler 10b (3.33 $M_\oplus$), Kepler 36b (4.45 $M_\oplus$), Kepler 78b (1.87 $M_\oplus$), and Kepler 93b (3.80 $M_\oplus$) are exoplanets (Table 6) with a geophysical class $P4R_S$ similar to the terrestrial planets of the Solar System. As discussed in § 3.6, while these planets are 1.87-4.45 times more massive than the Earth, they fall within normal mass range for the population of rock planets and therefore these planets can be identified as “Earths” rather than “super-Earths”. Kepler 78b with a mass of 1.87 $M_\oplus$ and a density of 6.0 g $cm^{-3}$ most likely has an iron content of ~32% with the remaining mass as silicate rocks (Grunblatt, Howard, & Haywood 2015). The composition and internal structure of Kepler 78b therefore should be most similar to the Earth’s among the exoplanets listed in Table 6.

**Table 5**. Planetary Geophysical Classification Applied to Solar System Planetary Bodies

| Planet | Classification | Planet | Classification |
|---|---|---|---|
| Mercury | $P4R_M$ | Rhea | $P5I_S$ |
| Venus | $P4R_S$ | Titan | $P5R_I$ |
| Earth | $P4R_S$ | Iapetus | $P5I_S$ |
| Moon | $P5R_S$ | Uranus | $P3I_{SG}$ |
| Mars | $P4R_S$ | Miranda | $P5I_S$ |
| Ceres | $P5R_I$ | Ariel | $P5I_S$ |
| Pallas | $P5R_I$ | Umbriel | $P5I_S$ |
| Vesta | $P5R_S$ | Titania | $P5I_S$ |
| Jupiter | $P2G_{01}$ | Oberon | $P5I_S$ |
| Io | $P5R_S$ | Neptune | $P3I_{SG}$ |
| Europa | $P5R_I$ | Triton | $P5R_I$ |
| Ganymede | $P5R_I$ | Proteus | $P5I_S$ |
| Callisto | $P5I_S$ | Pluto | $P5R_I$ |
| Saturn | $P2G_{23}$ | Charon | $P5R_I$ |
| Mimas | $P5I_S$ | Eris | $P5R_I$ |
| Enceladus | $P5R_I$ | Makemake | $P5R_I$ |
| Tethys | P5I | Haumea | $P5R_I$ |
| Dione | $P5I_S$? | | |

**Table 6**. Planetary Geophysical Classification Applied to a Sample of Exoplanets

| Planet | Possible Classification | Reference |
|---|---|---|
| Kepler 10b | $P4R_S$ | Dumusque et al. 2014 |
| Kepler 10c | $P3R_I$ - $P3R_S$ | Dumusque et al. 2014 |
| Kepler 77b | $P2G_{02}$ | Gandolfi et al. 2013 |
| Kepler 36b | $P4R_S$ | Carter et al. 2012 |
| Kepler 36c | $P3R_G$ | Carter et al. 2012 |
| Kepler 11b | $P4R_G$ – $P4I_S$ | Lissauer et al. 2013 |
| Kepler 11c | $P4R_G$ | Lissauer et al. 2013 |
| Kepler 11d | $P3R_G$ | Lissauer et al. 2013 |
| Kepler 11e | $P3R_G$ | Lissauer et al. 2013 |
| Kepler 11f | $P4R_G$ | Lissauer et al. 2013 |
| HD 149026b | $P2I_G$ – $P2I_{SG}$ | Fortney et al. 2006 |
| GJ 436b | $P3I_{SG}$ - $P3R_{IG}$ | Gillon et al. 2007; Nettelmann et al. 2010 |
| Kepler 30d | $P3G_{24}$ | Spiegel, Fortney, & Sotin 2014 |
| Kepler 68b | $P3R_I$ - $P3R_G$ – $P3I_S$ | Gilliland et al. 2013 |
| Kepler 78b | $P4R_S$ | Grunblatt, Howard, & Haywood 2015 |
| CoRoT-7b | $P3R_M$ – $P3R_S$ | Wagner et al. 2012, Barros et al. 2014 |
| HD 97658b | $P3R_I$ - $P3R_G$ - $P3R_{IG}$ | Van Grootel et al. 2014 |
| KOI-423b | $P1G_D$ - $P1G_{01}$ | Bouchy et al. 2011 |
| Kepler 18b | $P3R_I$ – $P3I_S$ | Cochran et al. 2011 |
| Kepler 18c | $P3I_{SG}$ | Cochran et al. 2011 |
| Kepler 18d | $P3I_{SG}$ | Cochran et al. 2011 |
| GJ 3470b | $P3R_G$ - $P3I_{SG}$ | Demory et al. 2013 |
| Kepler 20b | $P3R_G$ - $P3I_S$ | Gautier III et al. 2012 |
| Kepler 20c | $P3R_G$ – $P3I_S$ | Gautier III et al. 2012 |
| Kepler 454b | $P4R_S$ | Gettel et al. 2016 |
| Kepler 93b | $P4R_S$ | Ballard et al. 2014; Dressing et al. 2015 |
| BD+20594b | $P3R_S$ – $P3R_I$ | Espinoza et al. 2016 |
| KOI 614b | $P2G_{02}$ | Almenara et al. 2015 |
| KOI 680b | $P2G_{01}$ | Almenara et al. 2015 |
| KOI 206b | $P2G_?$ | Almenara et al. 2015 |
| Kepler 101b | $P2I_{SG}$ | Bonomo et al. 2014 |
| Kepler 423b | $P2G_{01}$ | Gandolfi et al. 2015 |
| Wasp 46b | $P2G_{01}$ | Ciceri et al. 2016 |
| Wasp 45b | $P2G_{12}$ | Ciceri et al. 2016 |
| HATS-17b | $P2G_{25}$ | Brahm et al. 2016 |

Gas giant planets in the P2 mass class vary in composition with clues to their structure found from mass-density and mass-radius relationships. For example, the planet KOI 206b's density is 1.13 g cm$^{-3}$ but its mass is 2.82 $M_J$ implying that the planet must have a different internal structure than Jupiter (Almenara et al. 2015) and is therefore a $P2G_{05}$ planet. KOI 614b (2.86 $M_J$) can have at most 16.5% by mass Z>2 elements (Almenara et al. 2015) and therefore has a $P2G_{02}$ classification where "$G_{02}$" indicates a gas planet with between 0 and 20% Z>2 elements by mass.

Planetary bodies formed in a proto-planetary disk that exceed the deuterium burning limit are 'deuterium burning planets' (Baraffe et al. 2014) and are classified as $P1G_D$ planets in the taxonomy presented here. A possible candidate for this planet class is KOI-423b which is an 18 $M_J$ body that is either a GP or a BD (Bouchy et al. 2011).

## 5. Conclusion

As the list of known exoplanets grows and the availability and quality of data for modeling the composition and interior structure of exoplanets improves a system of codes is needed to aid in quick comparisons of basic planetary geophysical characteristics. The system of codes presented in this paper provides a simple geophysical taxonomy that addresses this need. The composition codes are flexible enough to accommodate new planet classes identified in exoplanet studies and when combined with the mass scale codes allow for useful geophysical comparisons between exoplanets and the spherical planetary bodies of the Solar System.

The mass and composition codes presented in this paper can serve as a foundation for a more extensive system of codes that might include additional physical characteristics such as planetary temperature or density and dynamical characteristics such as orbital semi-major axis. The system may also be combined with other systems of codes already proposed. For example, the taxonomy of Irwin & Schulze-Makuch (2001) uses roman numerals I-V to rank the probability of life for a planet. If the Irwin & Schulze-Makuch taxonomy is combined with the taxonomy proposed in this paper then the Earth would be coded as a $P4R_SI$, Europa would be coded as a $P5R_III$, and Jupiter would be coded as a $P2G_{01}V$.

## Appendix: Dynamical Classes for Spherical Planetary Bodies

The planetary mass scale and the composition classes described in this paper provide a geophysical classification system that can be applied to spherical sub-stellar mass bodies irrespective of their dynamical status as planets, dwarf planets, or moons. For example Ganymede, Pluto, and Kepler 18b are classified as a moon, a dwarf planet, and a planet respectively but all have composition class "$R_I$". Therefore a dynamical code can be a valuable addition to the geophysical classification system described in this paper. Dynamical codes are provided in Table A1.

The definitions that follow describe the general dynamical circumstances for spherical planetary bodies and correspond with the dynamical codes (Table A1). These codes may be added to the geophysical classification described in Section 4. The term "***planetary body***" is a general term for spherical sub-stellar mass bodies including planets, dwarf planets, and spherical moons.

**Table A1**. Dynamical Class Codes for Spherical Planetary Bodies

| Dynamical Class | Dynamical code |
|---|---|
| Principal planet (Planet) | p |
| Belt planet (dwarf planet) | b |
| Moon | m |
| Rogue planet | r |
| Principal double planet | pd |
| Belt double planet | bd |
| Rogue double planet | rd |

**Principle planet**: A planetary body orbiting a star or brown dwarf that is dynamically dominant within its orbital zone. A Principal planet will have significantly more mass than the cumulative mass of all other bodies sharing in the Principal planet's orbital zone.

**Belt planet:** A planetary body orbiting a star or brown dwarf that does not dynamically dominate its orbital zone. A Belt planet shares its orbital region with numerous mostly smaller bodies having a cumulative mass larger than the mass of the belt planet.

**Moon:** A planetary body orbiting a larger planetary body and with the orbital barycenter inside the radius of the larger body.

**Double planet or double belt planet:** A pair of Principal or Belt planets with an orbital barycenter outside the radius of the larger planetary body.

**Rogue Planet:** A free-floating planetary body ejected from a planetary system which does not orbit any star or brown dwarf.

**Satellite:** A satellite is not a type of planetary body but is a sub-planetary mass ($< \sim 3 \times 10^{19}$ kg) body orbiting a Principal planet, Belt planet, Moon, or Rogue planet.

In common usage the terms "moon" and "satellite" are generally interchangeable terms. However, in the system of dynamical classes described above the two terms have distinct meanings. Non-spherical sub-planetary mass bodies orbiting planets and dwarf planets are "satellites" and considered a separate class from the spherical "moons".

With regard to different definitions for the term "planet" it is important to note that the dynamical classes described above are compatible with the International Astronomical Union (IAU) planet definition and also compatible with a geophysical definition in which all spherical planetary bodies are classified as "planets" (e.g. Stern & Levison 2002, Runyon et al. 2017).

If the IAU's planet definition is preferred then the dynamical classes "Principal planet" and "Belt planet" align with the IAU definitions for "planet" and "dwarf planet" respectively. Bodies identified as "belt planets" and the spherical "moons" are not planets in the IAU definition. However all planets, dwarf planets, and spherical moons can be assigned a mass class code from the planetary mass scale (Table 3) and a composition class code (Table 1). The dynamical class codes can also be assigned although sticking strictly to the IAU definitions it would be more appropriate to identify, for example, the Earth, Pluto, and Ganymede as a "$P4R_S$ planet", a "$P5R_I$ dwarf planet", and a "$P5R_I$ moon" respectively.

If the geophysical definition of "planet" is preferred then the dynamical classes "Principal planet", "Belt planet", and "moon" are dynamical classes of planets. Each planet would have a mass class code from the planetary mass scale (Table 3), a composition class code (Table 1) and a dynamical class code (Table A1).

Whether the IAU planet definition or the geophysical planet definition is preferred, any given spherical planetary body will be assigned the exact same mass scale code, composition code, and dynamical code as those codes have been defined in this paper. The classification for Solar System planetary bodies when combining the dynamical code with the geophysical codes is provided in Table A2.

**Table A2:** Classification of Solar System Planetary Bodies

| Solar System planetary body | Classification |
|---|---|
| Mercury | $P4R_{M}p$ |
| Venus | $P4R_{S}p$ |
| Earth | $P4R_{S}p$ |
| Moon | $P5R_{S}m$ |
| Mars | $P4R_{S}p$ |
| Ceres | $P5R_{I}b$ |
| Jupiter | $P2G_{01}p$ |
| Saturn | $P2G_{23}p$ |
| Uranus | $P3I_{SG}p$ |
| Neptune | $P3I_{SG}p$ |
| Pluto | $P5R_{I}bd$ |
| Charon | $P5R_{I}bd$ |
| Io | $P5R_{S}m$ |
| Europa | $P5R_{I}m$ |
| Ganymede | $P5R_{I}m$ |
| Callisto | $P5I_{S}m$ |
| Titan | $P5R_{I}m$ |
| Iapetus | $P5I_{S}m$ |
| Mimas | $P5I_{S}m$ |
| Tethys | P5Im |
| Triton | $P5R_{I}m$ |

## Acknowledgements

This research has made use of NASA's Astrophysics Data System Bibliographic Services. This research has made use of the Exoplanet Orbit Database and the Exoplanet Data Explorer at exoplanets.org. I would like to thank David Merritt for helpful comments on the composition codes presented in Table 1.